\documentclass[a4paper,11pt]{article}
\usepackage{jinstpub} 
\usepackage{lineno}
\usepackage{textgreek}
\usepackage{booktabs}
\usepackage{svg}

\nolinenumbers

\title{\boldmath A novel liquid argon purity monitor  based on ${}^\mathsf{207}$Bi}

\author[a]{B. Baibussinov}
\author[a]{M. Bettini}
\author[a]{F. Fabris}
\author[e]{R. Gan}
\author[a]{A. Guglielmi}
\author[b,c]{G. Gurung}
\author[a]{S.~Marchini}
\author[a]{G. Meng}
\author[a]{M. Nicoletto}
\author[b,a]{F. Pietropaolo\note{Corresponding author.}}
\author[b]{X. Pons}
\author[a]{G. Rampazzo}
\author[a]{R. Triozzi}
\author[a]{F. Varanini}
\author[d]{and B. Voirin}

\affiliation[a]{Sezione INFN di Padova e Università di Padova, via Marzolo 8, 35131, Italy}
\affiliation[b]{CERN, Route de Meyrin, 1211 Geneva, Switzerland}
\affiliation[c]{Department of Physics, University of Texas, Arlington, TX 76019, USA}
\affiliation[d]{Department of Physics, École Normale Supérieure de Lyon, 69364 Lyon Cedex 07, France}

\affiliation[e]{Department of Physics, Boston University, Boston, MA 02215, USA}

\emailAdd{francesco.pietropaolo@cern.ch}

\abstract{A novel liquid argon purity monitor based on a ${}^{207}$Bi radioactive source, emitting monochromatic internal-conversion electrons, is presented.
This new monitor allows for a very precise and fast measurement of the electronegative impurities concentration in liquid argon. It can be operated continuously in liquid argon TPC experiments without interfering with the main detector operation.
Different drift lengths can be assembled for the proposed device, to assess a large range of liquid argon purities while minimizing systematic uncertainties.
Two prototypes have been built and successfully operated in dedicated test stands. The results and performance are reported.
}


\begin{document}
\maketitle
\flushbottom

\section{Introduction}
\label{sec:intro}

Liquid Argon (LAr) has emerged as the preferred material target for numerous current and future experiments aimed at detecting neutrinos and exploring unexpected phenomena in astro-particle physics. Notably, the Liquid Argon Time Projection Chamber (LArTPC) technique, introduced by C. Rubbia in 1977 \cite{rubbia1977}, provides high-resolution imaging of charged particles on the millimeter scale by collecting ionization electrons that drift toward the anode wire planes in a uniform electric field \cite{icarus}.

The collected charge signal is subject to attenuation due to the recombination of ionization electrons and argon ions, which depends on the applied drift field; at a typical 500~V/cm electric field the recombination is about one-third of the initial ionization charge \cite{recombination}. 

Additionally, the signal can be further attenuated by the absorption of the ionization electrons by electro-negative impurities such as O$_2$ present in the liquid argon. The related attenuation is described by an exponential decay function, $\sim\exp(-t_D / \tau_e)$, where $t_D$ is the drift time and $\tau_e$ is the free electron lifetime.
In currently operating LArTPCs, free electron lifetimes exceeding $\sim10 \ \text{ms}$ (corresponding to $\sim 30$ p.p.t. of O$_2$ residual impurities) are routinely achieved \cite{icarino-lifetime, icarus-lifetime, protodune}, resulting in a signal attenuation of $\sim 6\%$ per meter of drift for a $500 \ \text{V/cm}$ field ($v_D \sim 1.5 \ \text{mm/μs}$).

Monitoring liquid argon purity during LArTPC operation is crucial, and the associated electron lifetime can be assessed by measuring signal attenuation along cosmic muon tracks that span the entire drift volume \cite{icarus-lifetime}. However, for deep-underground experiments, cosmic muon signals are rare and only available when the detector is fully filled and operational. To address this issue, a specialized UV-based liquid argon purity monitor was developed by G. Carugno et al. during the early phases of the ICARUS LArTPC research and development \cite{carugno-uv-pm} and is presently utilized by most of the operational LArTPCs. 

An alternative approach based on a ${}^{207}$Bi radioactive source along with a new, better performing setup is proposed.

\section{The new~\boldmath{$^{207}$}Bi LAr purity monitor}
\label{sec:pm}

The most widely used LAr purity monitors are double-gridded TPCs where an electron cloud is generated on the cathode by photoelectric effect from a Xenon lamp UV light flashes, routed to the cathode by a fiber bundle.
The electron signal attenuation is given by the ratio of charge reaching the anode ($Q_A$) and the one leaving the cathode ($Q_C$), measured at each light flash. The corresponding electron lifetime is determined as $Q_A / Q_C = \exp(-t_D / \tau_e)$.

\begin{figure}[htbp]
\centering
\includegraphics[width=1\textwidth]{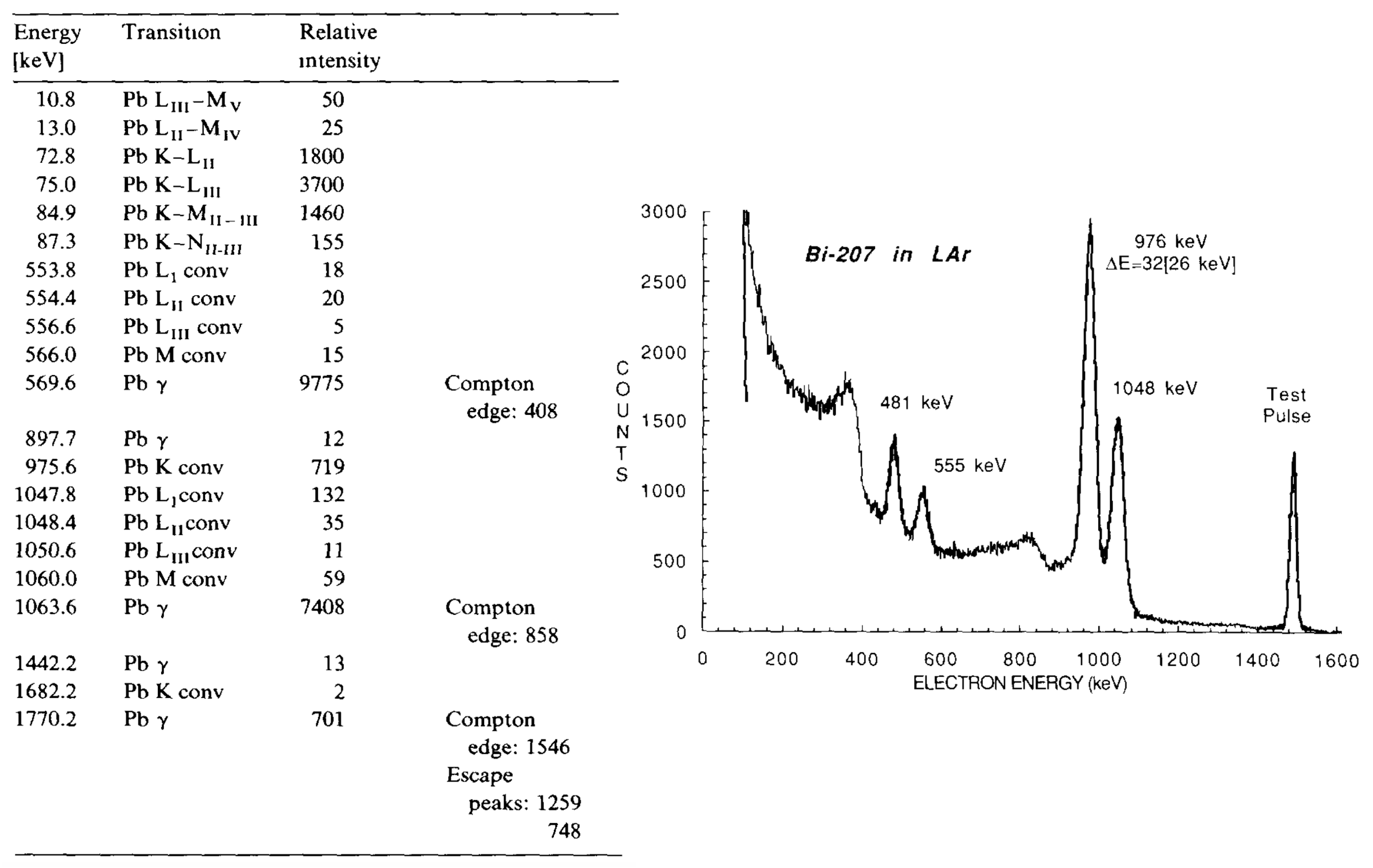}
\caption{(Left) Transitions and their relative intensities of ${}^{207}$Bi. (Right) Pulse height spectrum of the ${}^{207}$Bi radioactive source in liquid argon at a drift field of 10.9 kV/cm. The width of the IC peaks is dominated by the electronic noise as evaluated with the test-pulse signal also included in the spectrum at $\sim1500 \ \text{keV}$. 
Both the transition table and the spectrum are from \cite{aprile-bi207}.}
\label{fig:bi207}
\end{figure}

A novel purity monitor concept based on a ${}^{207}$Bi radioactive source, which decays to ${}^{207}$Pb by emitting monochromatic internal-conversion (IC) electrons primarily at $\sim976 \ \text{keV}$ and at $\sim1048 \ \text{keV}$, is proposed. The excited ${}^{207}$Pb can transition to its ground state by emitting gamma rays or IC electrons up to 1.7 MeV. The gamma rays at approximately 1063 keV generate Compton electrons, producing a Compton edge at $\sim860 \ \text{keV}$, which is energetically close to the two IC electron signals around $\sim 1 \ \text{MeV}$ (see figure \ref{fig:bi207}).

This source is frequently used in ultrapure cryogenic environments and is employed in noble liquid-based experiments for calibration purposes with IC electrons. Historically, ${}^{207}$Bi has been used to investigate the energy resolution of liquid argon calorimeters and to measure liquid argon purity at the parts per million (ppm) level \cite{aprile-bi207, adams-bi207}. 

\begin{figure}[htbp]
\centering
\includegraphics[width=1\textwidth]{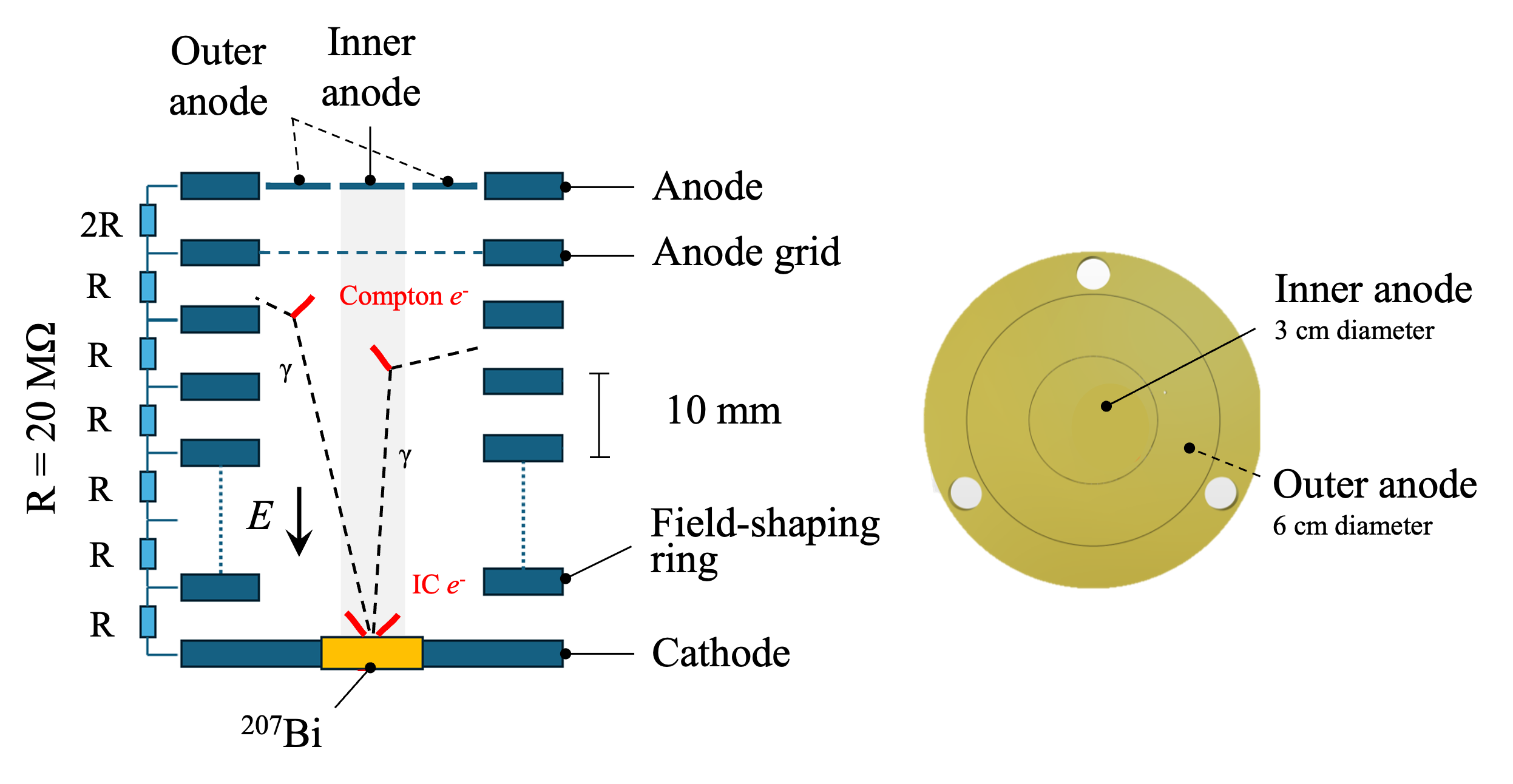}
\caption{(Left) Schematic of the ${}^{207}$Bi-based purity monitor concept. The 5 mm source is embedded in the cathode. The inner anode receives IC and Compton electrons, the outer anode only collects Compton electrons. 
(Right) Detail of the anode, split into three concentric 3, 6 and 8 cm diameter areas.\label{fig:pm}}
\end{figure}

The design of the new purity monitor (PM) is derived from the previously described UV-monitors, with some innovative concepts (figure \ref{fig:pm}). It consists of a cylindrical drift chamber with an active LAr volume diameter of $6 \ \text{cm}$ and a drift length of the order of $10 \ \text{cm}$.

A ${}^{207}$Bi source with a $5 \ \text{mm}$ diameter is embedded in the cathode whose diameter is $8 \ \text{cm}$. The field cage rings have an outer and inner diameters of 8 and $6 \ \text{cm}$ respectively. The anode is split into three concentric areas with diameters of 3, 6 and 8 cm respectively:
\begin{itemize}
    \item the inner anode receives the ionization electron cloud generated by the IC electrons within a few mm from the source and drifted in liquid argon by the uniform electric field in the PM; it also collects the ionization signal of the Compton electrons produced by $\gamma$s emitted from the source and interacting in the innermost 3 cm diameter PM volume;
    \item the outer anode only receives signals from Compton electrons produced in the outermost PM volume with 3 to 6 cm radius; the energy spectra of the Compton electrons collected on the inner and outer anodes are similar, therefore the outer energy spectrum can be used to remove the Compton electron background from the inner spectrum, hence extracting the pure IC electron peak.
    \item the most external anulus acts as the last grounded electrode of the field cage.
\end{itemize}

The proposed new-concept purity monitor presents several advantages:
\begin{itemize}
    \item the ${}^{207}$Bi sources are available with activities up to 400 kBq; its half-lifetime is of 31 years, well in excess of the designed time exposures of future detectors (e.g., DUNE \cite{dune});
    \item the energy of the IC electrons is similar to a typical deposition of a minimum ionizing particle in liquid argon collected on a $5\ \text{mm}$ anode readout channel; therefore a cryogenic front-end electronics similar to the one of the LArTPC could be used to readout the charge signals;
    \item no cathode readout is required, since the IC electron energy is known;
    \item it can be operated at the same drift field as the main LArTPC for a direct electron lifetime measurement; 
    \item PMs with different drift lengths can be assembled to measure the liquid argon purity in an extended range of values, further reducing the systematic uncertainty;
    \item the PM can be continuously operated without interfering with the photon detection system of the main LArTPC and also during the liquid argon filling phase.
\end{itemize}
Concerning the last item, it is worth mentioning that the Bi207 PM has been proposed to be used in experiments which address physics phenomena above the MeV energy scale of the gamma rays emitted by the ${}^{207}$Bi source.

\subsection{Design and construction of the monitors}

\begin{figure}[htbp]
\centering
\includegraphics[width=0.9\textwidth]{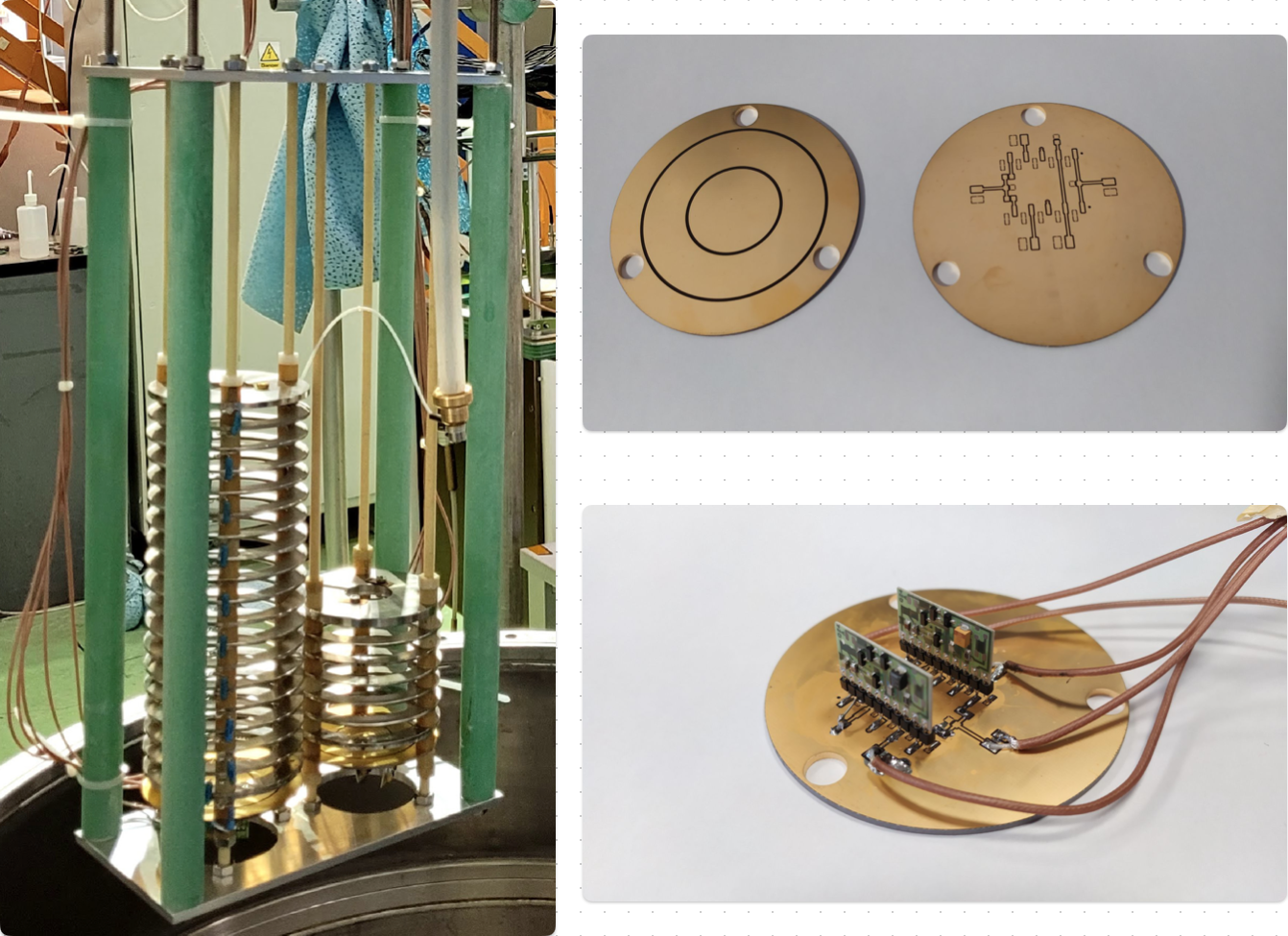}
\caption{(Left) The two purity monitor prototypes with 6 and 18 cm drift lengths. (Right) Detail of the anodes, with the pre-amplifiers mounted on their back. \label{fig:pm-prototype}}
\end{figure}

Two purity monitors prototypes have been built at CERN and INFN Padova with exactly the same layout, except for the drift lengths of 6 and 18 cm for the ``short'' and ``long'' devices (figure \ref{fig:pm-prototype}). 
Two $\sim 37 \ \text{kBq}$ ${}^{207}$Bi radioactive sources, protected by titanium thin foils leading to a $< 4 \ \text{keV}$ energy loss for IC electrons, are embedded into the cathodes of both PMs.

The high voltage is applied on the cathode of the long purity monitor. The chain of 20 MΩ resistors applied to the field cage electrodes ensures a proper uniformity of the electric field, as described in \cite{carugno-uv-pm}. The short PM is positioned beside the long one with the anode and the field cage electrodes at the same level as those of the long one. This allows to derive the voltages of the cathode and the field cage electrodes of the short PM from the electrodes of the long one. 
This configuration ensures that the electric field in the long and short PMs is exactly the same allowing, when the two PM are used simultaneously, to remove absolute calibration requirements. 
This is beneficial for the measurement of high electron lifetimes, where the main systematic uncertainty is due to the cross-calibration of the readout electronics in the two devices. 
Lower electron lifetimes can be directly measured by using the short purity monitor in stand-alone mode.

High voltage feed-through and cables, developed at CERN, allow to apply voltages $>20 \ \text{kV}$ and operate the purity monitors at a drift field up to 1 kV/cm, in the range of those of LArTPC experiments, providing a direct estimation of the liquid argon purity at the nominal drift field.

\begin{figure}[htbp]
\centering
\includegraphics[width=0.8\textwidth]{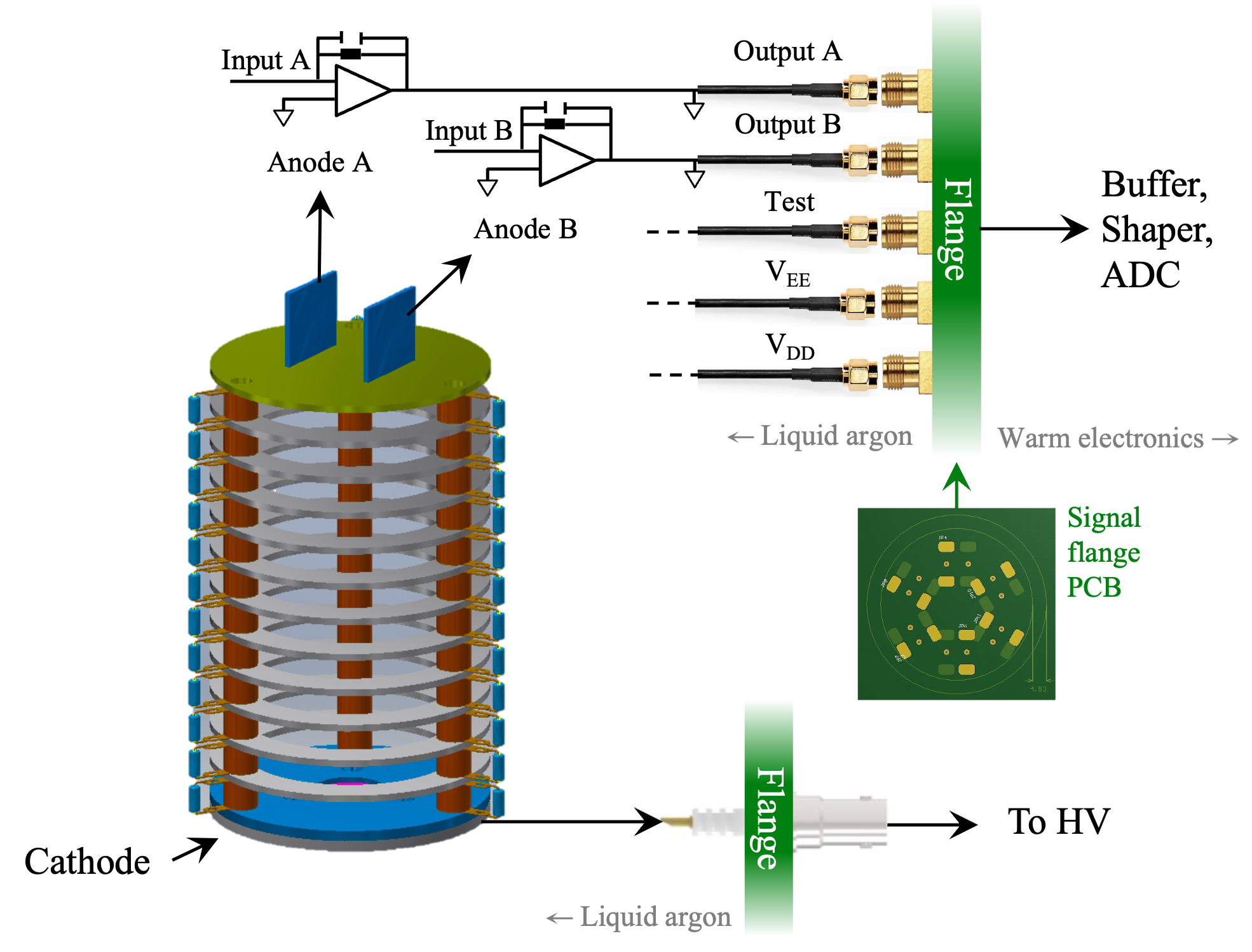}
\caption{General layout of the electronics for the ${}^{207}$Bi purity monitors: the pre-amplifiers are mounted on the back of the two anodes in the liquid argon to minimize the detector input capacitance and the related electronics noise.  \label{fig:electronics}}
\end{figure}

The readout electronics consists of pre-amplifiers able to operate in LAr directly mounted on the back of the anodes, coupled to warm buffer amplifiers and shapers installed on the signal feed-through flange outside the liquid argon cryostat (figure \ref{fig:electronics}).

The charge-sensitive pre-amplifiers are based on the jFET TOTEM architecture \cite{icarus-totem} (table \ref{table:totem}), originally developed for cryogenic operations in the ICARUS experiment. They are characterized by: low-power, $\sim 500 \ e^-$ low-noise, ability to drive $> 15 \ \text{m}$-long output cable with no loss, and high reliability to sustain several cryogenic cycles for $>30$ years.
\begin{table}[ht]
\centering
\begin{tabular}{lc}
\toprule
Sensitivity & $\sim0.45 \ \text{mv/fC}$ ($\sim0.9 \ \text{μA/fC}$) \\
Dynamic range & $\pm 1.5 \ \text{pC}$\\
Linearity & < 0.5\% at full scale\\
Input impedance & $\sim420 \ \text{Ω}$\\
Input capacitance & $\sim20 \ \text{pF}$ \\
E.N.C. & $\sim(390 + 7 \times C_D/pF)$ $e^-$ \\
Power consumption & $\sim11 \ \text{mW}$ \\
\bottomrule
\end{tabular}
\caption{Characteristics of the charge sensitive pre-amplifier based on the TOTEM architecture \cite{icarus-totem}.}
\label{table:totem}
\end{table}
For each PM, the pre-amplifiers are selected with a similar response, gain, and calibration capacitance with a better than $2\%$ accuracy. Also, no decoupling capacitors are required, as the anodes are set at a tension of $0 \ \text{V}$. The signal feed-through flanges, warm receivers and signal shapers were designed and built at INFN Padova. The warm buffer amplifier, based on a high-performance AD811 operational amplifier, drives the signal from the pre-amplifier to the shaper. Stable bias voltages to the pre-amplifiers are provided by DC-DC converters included in the external warm electronics.

\begin{figure}[htbp]
\centering
\includegraphics[width=0.6\textwidth]{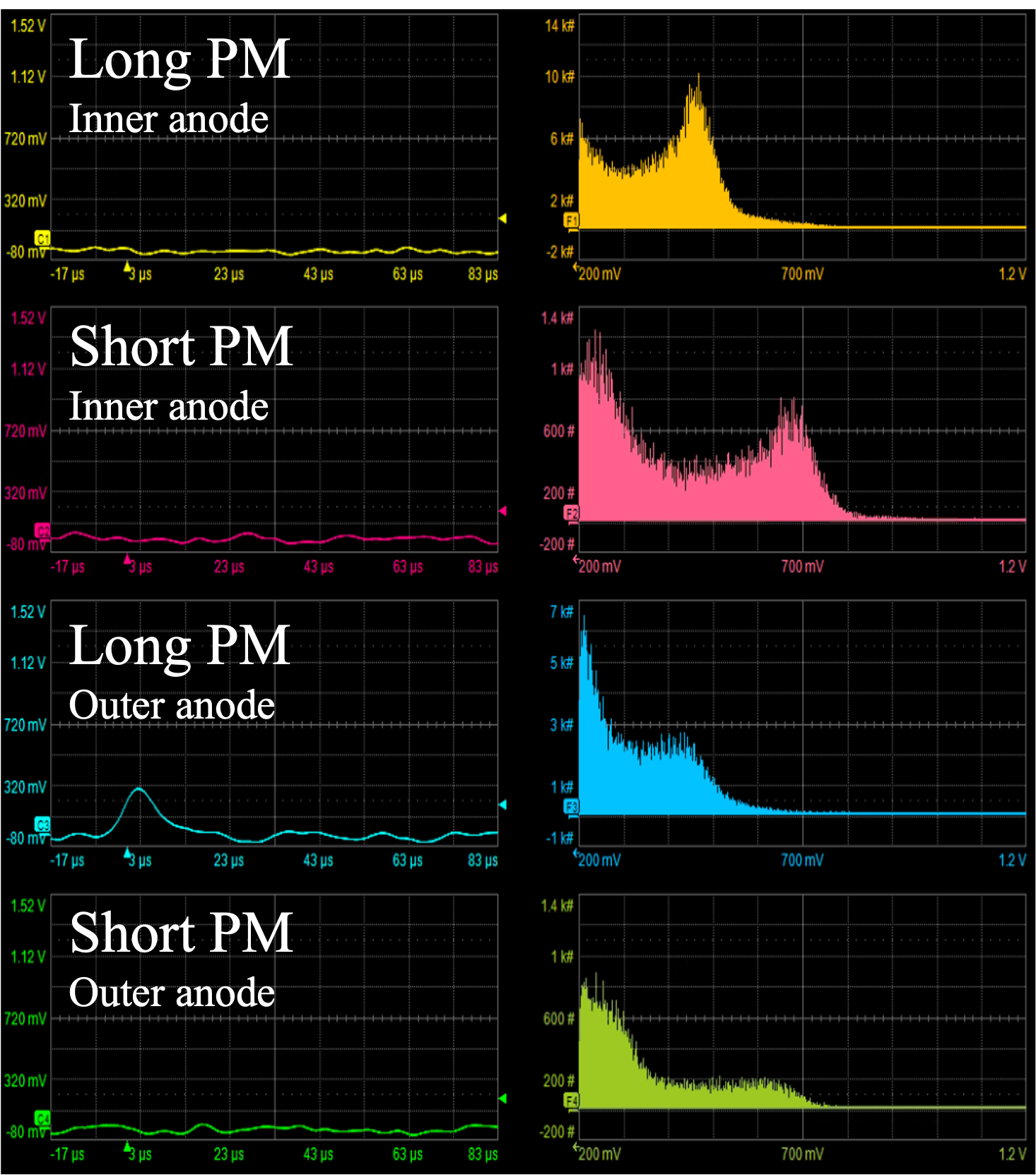}
\caption{${}^{207}$Bi spectra acquired with the oscilloscope-based readout system. The trigger is set as the OR of the four channels, which correspond to the inner and outer anodes of the two PM. The left panels show single anode waveforms, while the collected spectra are shown on the right. In the inner spectra, the IC electron peaks are visible on top of the Compton electron background. \label{fig:oscilloscopes}}
\end{figure}

An oscilloscope-based DAQ system readouts and stores the signal wave-forms for off-line analysis of charge measurements through either peak sensing or area calculation (figure \ref{fig:oscilloscopes}). 

In the future, a multi-channel analyzer system, relying on peak sensing with considerably less dead-time, is expected to replace the oscilloscope system.

\section{Liquid argon purity measurement}
\label{sec:results}

\begin{figure}[htbp]
\centering
\includegraphics[width=0.99\textwidth]{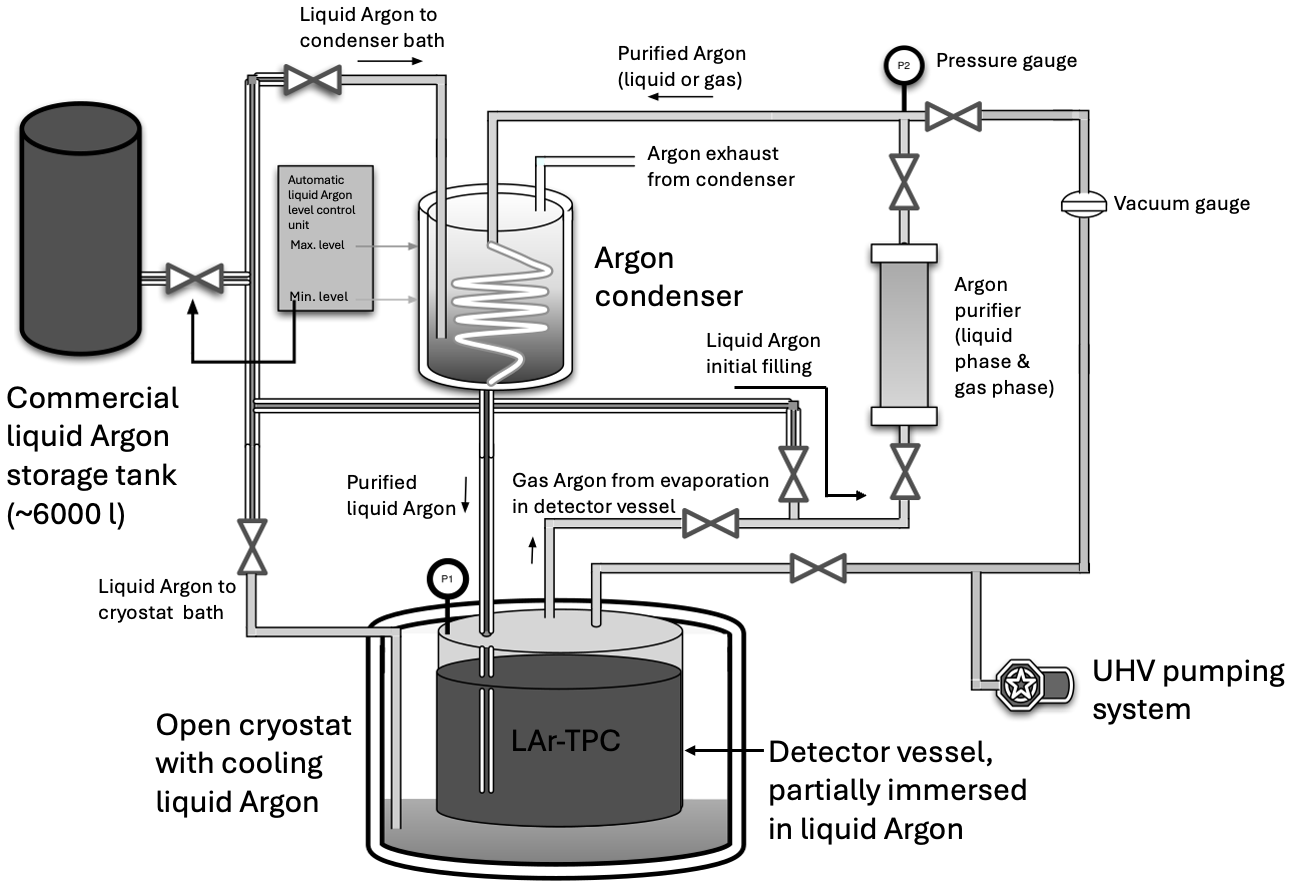}
\caption{Schematics of the cryogenic setup at CERN used for the PMs tests. \label{fig:LAr_recirculation}}
\end{figure}

Extensive tests of the developed purity monitors were performed at CERN in a dedicated vessel, containing up to 250 kg of LAr. The vessel is equipped with a LAr purification / recirculation system, and it is inserted in an open-air cryostat filled with commercial LAr acting as cryogenic bath (figure \ref{fig:LAr_recirculation}).
The PMs vessel is initially evacuated down to $\sim 10^{-5}$ mBar with turbo-molecular pumping system, to allow degassing of the detector materials, then cooled down in the LAr bath and finally filled with commercial LAr purified (in liquid phase) by a molecular sieve to remove H$_2$O and CO$_2$ followed by a Copper based filter to remove O$_2$. Initial impurity concentrations better than $\sim$ ppb of O$_2$ equivalent are routinely reached.
During the operation, the gaseous Argon evaporating from the liquid in the PMs vessel is continuously purified by the same filters, condensed in the passive heat exchanger (LAr bath) and re-injected into the detector vessel. With a recirculation flow of 3 to 4 kg/h, the entire LAr volume is recirculated in less than three days, allowing to reach a steady impurity level better than 0.1 ppb of O$_2$ equivalent in about a week.
Cryogenic systems with similar layout and performance are described in more details in \cite{icarino-lifetime, icarus-cryo}.

A quick evaluation of the liquid argon purity can be extracted by comparing the relative IC electron peak positions in the short and long PMs, where the systematic uncertainties are minimized by using the same layout.
On each PM, the inner and outer spectra are normalized in the region where only the main more intense Compton edge is present; the outer spectrum is then subtracted from the inner one to extract the IC electron signal.
This direct data-driven normalization procedure allows us to bypass indirect calculations related to the PM geometry, the DAQ system features such as the trigger conditions, and the peak identification methods. The scaling factors range from 1.2 to 2.0, depending on the PM drift length (6 or 18 cm), the applied electric field (500 to 900 V/cm) and the DAQ acquisition time window (100 to 200 us).

\begin{figure}[htbp]
\centering
\includegraphics[width=1\textwidth]{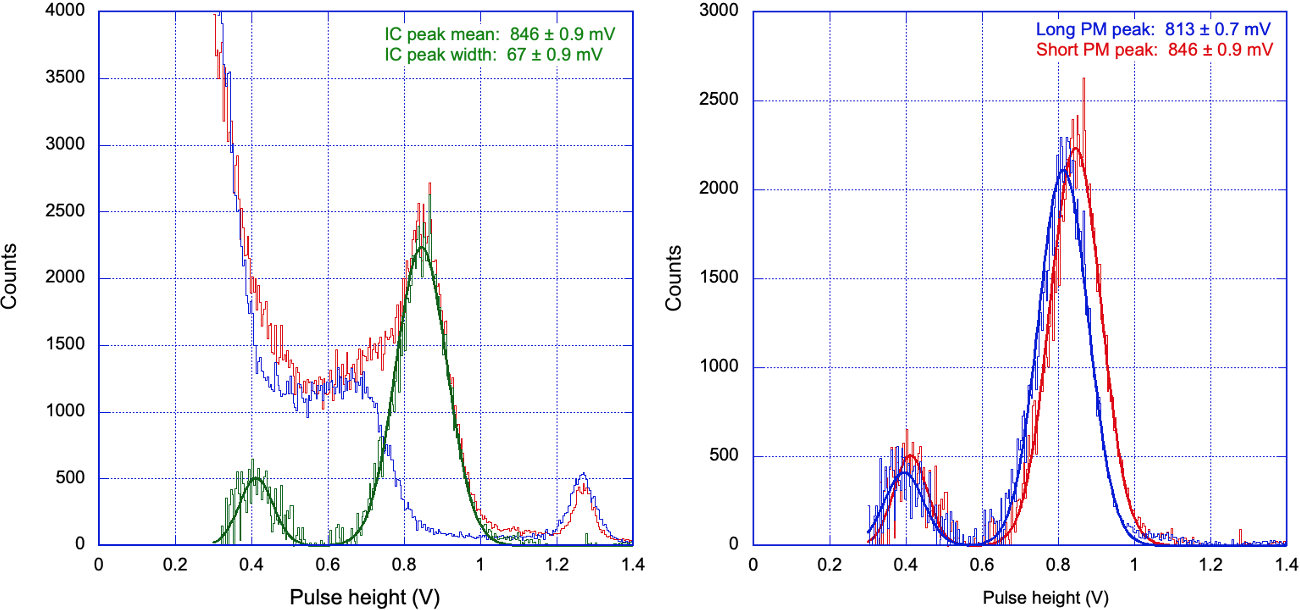}
\caption{(Left)  ${}^{207}$Bi pulse-height spectra acquired with the 60 mm long purity monitor. The inner and the outer anode spectra are shown in red and blue respectively. The outer spectrum is normalized to inner one where only the main Compton edge is present (the $0.5 \text{--} 0.7$ V range in the above example). The subtraction of the normalized outer spectrum from the inner one yields the clean $\sim 976 \ \text{keV}$ IC electron peak shown in green. The second IC electron peak at $\sim 500 \ \text{keV}$ is also partially visible after the subtraction. Test pulse signals are included between 1.2 and 1.4 V. (Right) IC electron peaks obtained with the short (60 mm drift) and long (180 mm drift) purity monitors. The ratio of the pulse heights is used to measure the electron lifetime. The spectra shown in this picture were taken with a drift electric field of 600 V/cm in both the long and the short purity monitors. \label{fig:spectra}}
\end{figure}

The resulting peak is fitted with a double Gaussian function to account for the two $\sim976$ and $\sim1048 \ \text{keV}$ IC lines, which are merged due to the $\sim 50 \ \text{keV}$ energy resolution, routinely achieved at drift fields in the $200 - 1000 \ \text{V/cm}$ range.
A statistical precision on the average energy of $\sim 0.1\%$ is reached after a few minutes of data acquisition (figure \ref{fig:spectra}).


After the IC electron peaks are extracted, the electron lifetime $\tau _{e}$ is evaluated as a function of the signal attenuation $A$:
\begin{equation*}
  A=\frac{Q_{L}}{Q_{S}} = \exp\left(\frac{t_{S}-t_{L}}{\tau _{e}}\right) \ 
\end{equation*}
where $Q_{L}$ and $Q_{S}$ are the average pulse heights in the long and short PM, respectively, and $t_L$ and $t_S$ are the corresponding drift times at the applied drift field.
The systematic uncertainty on the charge attenuation measured with the two PMs $\delta A / A$ is dominated by the $\sim1\%$ cross-calibrations of the pre-amplifiers, as determined by the test capacitance accuracy.

\begin{figure}[htbp]
\centering
\includegraphics[width=1\textwidth]{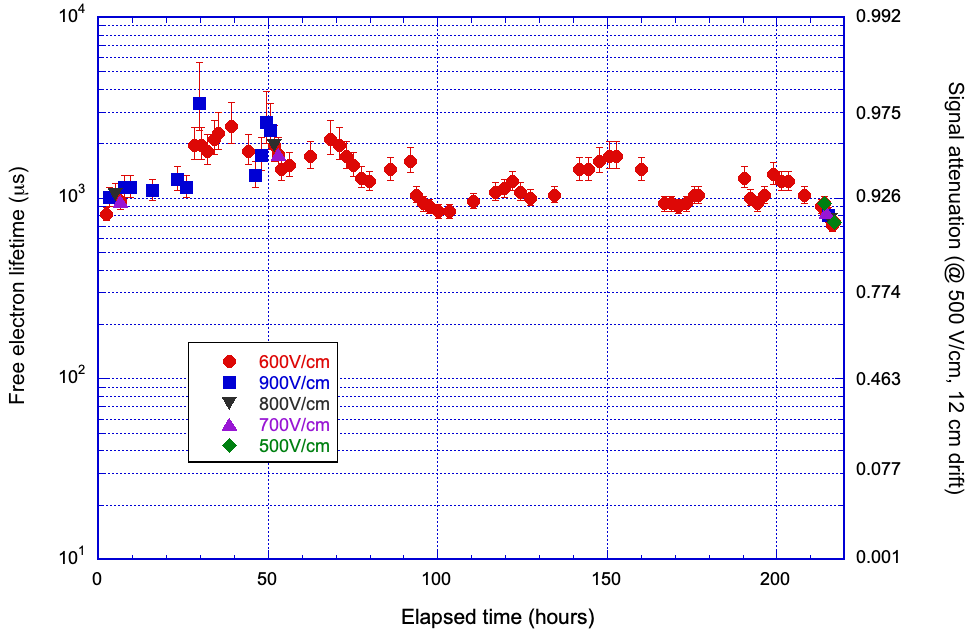}
\caption{Evolution of the electron lifetime as a function of time, obtained evaluating the ratio of the IC electron peak values in the long and the short PMs. Runs taken at different drift fields are shown in different colors, yielding compatible results. The equivalent signal attenuation between the short and the long purity monitors is shown on the right axis for the reference drift electric field of 500 V/cm. The modulation of the lifetime values are due to the varying LAr circulation rate in the test stand. After the test run, a small leak was found in the circulation system, most likely responsible of the negative trend of the LAr purity with time. \label{fig:lifetime}}
\end{figure}

For the spectra reported in figure \ref{fig:spectra}, acquired at a drift electric field of 600 V/cm, the signal attenuation $Q_{L}$ / $Q_{S}$ is $\sim$0.96 and the corresponding lifetime is $\sim$0.780 ms.
Figure \ref{fig:lifetime} shows the electron lifetime evolution as a function of time in one of the test runs. Measurements were performed at different electric fields, showing no significant dependency of the lifetime on the applied field.

The attenuation $A = Q_L/Q_S$ measured with the two PMs can be extrapolated to a longer drift path in the main LArTPC as 
\begin{equation*}
    A_{\text{TPC}} = A^{\Delta t_{\text{TPC}} / \Delta t}
\end{equation*}
where the $\Delta t_{\text{TPC}}$ and $\Delta t$ drift times refer to the LArTPC and PMs respectively.
The corresponding relative uncertainty is 
\begin{equation*}
   \frac{\delta A_{\text{TPC}}} {A_{\text{TPC}}} \sim \frac{\Delta t_{\text{TPC}}}{\Delta t} \; \frac{\delta A}{A} \ .
\end{equation*}
For a $1 \ \text{ms}$ free electron lifetime, the signal attenuation $A \sim 92.6\%$ is measured in the two $180 \ \text{mm}$ and $60 \ \text{mm}$ PMs at a drift field of 500 V/cm ($v_D = 1.5 \ \text{mm/μs}$) with $\sim1\%$ systematic accuracy, corresponding to $A_{\text{TPC}} \sim 51\% \pm 8\%$ when extrapolating $A$ to a $1 \ \text{m}$ drift length. 
It follows that the sensitivity of the present prototypes on the electron lifetime is estimated to be $3.3 \ \text{ms}$ at $90\%$ C.L.
A better sensitivity to larger electron lifetime values can be achieved by increasing the drift length difference or reducing the electric field in the two PMs.


An improved cross-calibration can be obtained by operating the PMs in a cryostat together with a LArTPC and comparing its result with electron lifetime measurements obtained evaluating the charge signal attenuation along cosmic muons tracks crossing the TPC drift volume.

Measurements with a single PM were performed, obtaining results in agreement with the ones from the twin monitors. 
For this purpose, the signal pulse height, $Q_0$, from the free electron charge produced at the cathode is required. $Q_0$ can be derived from the simultaneous measurements, $Q_S/Q_0 = \exp(-t_S/\tau_e)$ and $Q_L/Q_0 = \exp(-t_L/\tau_e)$, where $\tau_e$ is determined as described above and asuming that $Q_0$ is the same for both PMs within the systematic uncertainly due to the relative gain calibration and is independent from $\tau_e$.
High statistical accuracy, possibly better than the above mentioned $1\%$ systematics, can be achieved with multiple twin mode measurements. 
In this way, the short PM can be used to evaluate low electron lifetimes; increasing the drift electric field allows to further reduce the drift time, thus extending the explorable electrons lifetime range. The evaluation of $Q_0$ based on the twin PM operation, can be performed at all the drift fields planned for the single mode PM operation, to account for the different electron-ion recombination factors.
Notably, it was possible to detect the IC electron peak position for electron lifetimes down to tens of $\mu$s at a drift field of $1 \ \text{kV/cm}$.

To further validate the performance of the ${}^{207}$Bi PMs, a fast and simple Monte Carlo simulation has been set up \cite{MC-validation}. The layout of each purity monitor has been described as a cylindrical volume. Photons and electrons are randomly emitted from the ${}^{207}$Bi source, located on the cathode, according to their relative emission probabilities.
The interaction point for the photons is determined according to the scattering cross section; photoelectric  and pair production effects are not considered due to their negligible cross sections. If the interaction point occurs within the PM volume, the corresponding Compton electron energy is computed.

Because the track lengths of IC and Compton electrons do not exceed 4 mm in LAr, electrons are considered point-like in the simulation. For each electron, the generated deposited charge is reduced due to the electron-ion recombination and attenuated according to the electron drift path length and the pre-defined liquid argon purity.  The electron signal at the anode is smeared to account for the resolution of the charge-readout electronics.

\begin{figure}[htbp]
\centering
\includegraphics[width=\textwidth]{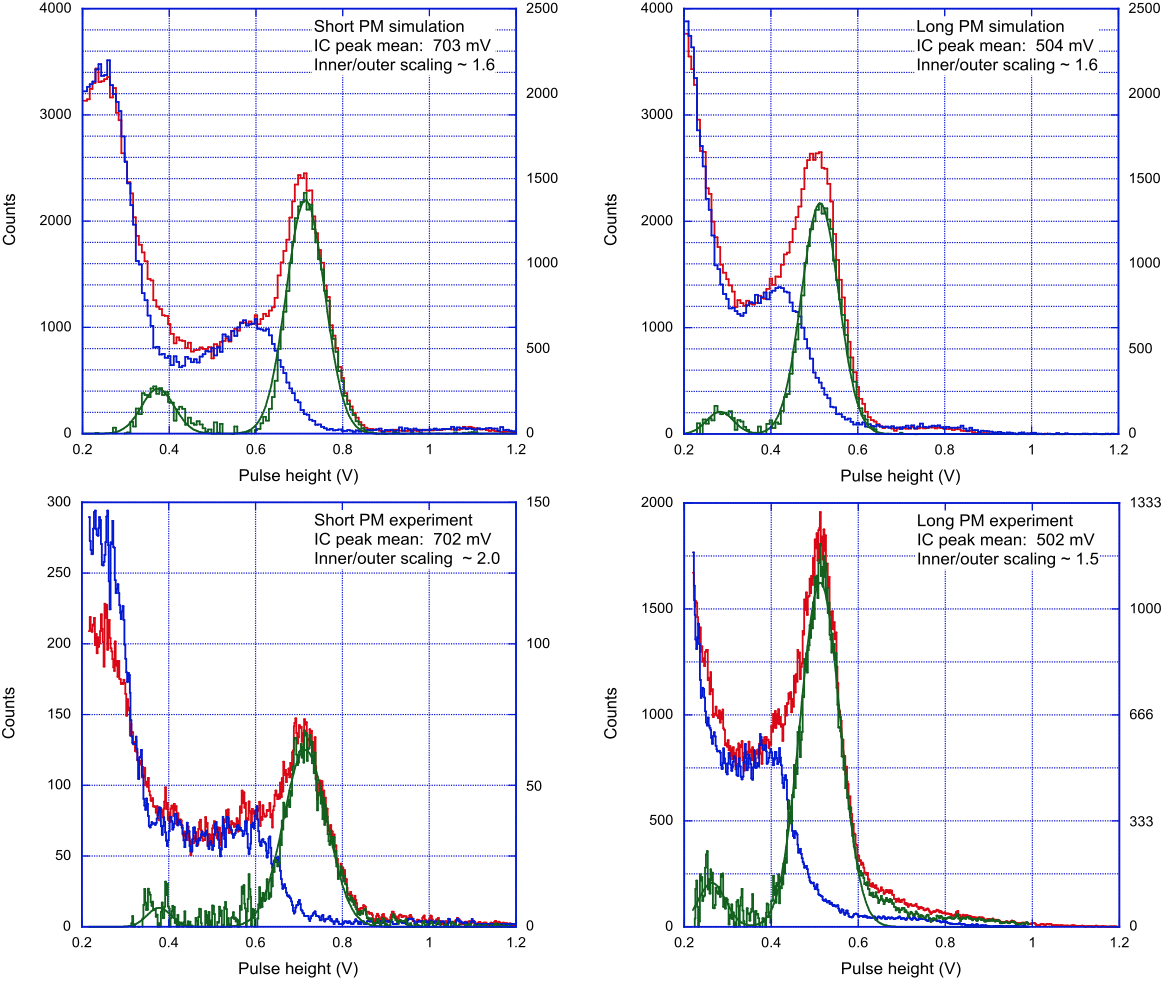}
\caption{Monte Carlo validation of the extraction of the IC electron peaks from the simulated short and long PMs. The free electron lifetime measured in the experimental data is applied into the Monte Carlo simulation. Data sets from a low purity ($\sim180 \ \text{μs}$) experimental run are shown to demonstrate that the energy spectra from both the long and short PMs reproduce well those derived from the experimental data over a large signal attenuation range.\label{fig:mc-validation}}
\end{figure}

The simulated data are analyzed in a similar way as the experimental ones: for both the long and the short PMs, the energy spectra are recorded. The spectrum on the outer anode is subtracted from the one on the inner anode and the resulting distribution is Gaussian-fitted to extract the mean energy and the width of the IC electron peak. Figure \ref{fig:mc-validation} shows the comparison of simulation and experiment for a data set with $\sim 180 \ \text{μs}$ electron lifetime. The agreement is highly satisfactory; this applies also to the whole purity range explored in the test runs.

\section{Conclusions}
\label{sec:conclusions}

A prototype of the new liquid argon purity monitor based on a ${}^{207}$Bi radioactive source has been designed, constructed and extensively tested to demonstrate its feasibility and to initially characterize the performance. 
LAr purities in the few tens of μs to several ms range have been assessed with two $18 \ \text{cm}$ and $6 \ \text{cm}$-drift length PMs with a $\sim1\%$ systematic accuracy on the signal attenuation. 
Improved sensitivities can be achieved by increasing the drift length difference or reducing the electric field in the two PMs.


The final validation will be performed by operating the PMs in the cryostat of a LArTPC, and comparing the PM results with those obtained evaluating the charge signal attenuation along muons tracks crossing the TPC drift volume.

The novel purity monitor presents several advantages, as previously discussed. In particular, it allows for a very quick high-statistics liquid argon purity measurement. 
The PM can be operated continuously without interfering with the main LArTPC operation and at the same electric field.
Different drift lengths can be assembled for the proposed device, to measure the liquid argon purity in a wide range of values while minimizing systematic uncertainties.


\acknowledgments

The support of CERN and INFN to the study and realization of a new liquid argon purity monitor based on the use of ${}^{207}$Bi is acknowledged.
The contribution of M. Meli to the realization of the mechanical components and the assembly of the purity monitor prototype is recognized and warmly acknowledged.




\end{document}